\documentclass[12pt]{article}
\usepackage{a4p}
\usepackage{epsfig}
\usepackage{array}
\usepackage{cite}
\usepackage{pennames}
\usepackage{rotating}

\newcommand{\etal}      {{\it et~al.}}
\newcommand{\PhysLett}  {Phys.~Lett.}
\newcommand{\PhysRep}   {Phys.~Rep.}
\newcommand{\PhysRev}   {Phys.~Rev.}
\newcommand{\PhysRevl}   {Phys.~Rev.~Lett.}
\newcommand{\NPhys}  {Nucl.~Phys.}

\newcommand{\CPC} {Comp.~Phys.\ Comm.}
\newcommand{\ZPhys}  {Z.~Phys.}

\newcommand{\ee}  {\mbox{$\rm e^+e^-\ $}}

\def\ee{\mbox{e}^+\mbox{e}^-}

\def\Z0{\mbox{Z}^0}

\begin{document}
\begin{titlepage}
\begin{center}
{\Large  EUROPEAN LABORATORY FOR PARTICLE PHYSICS} 
\end{center}
\bigskip\bigskip
\begin{flushright}
{\large  CERN-EP/99-091}\\ 
 7 July 1999
\end{flushright}
\bigskip\bigskip
\begin{center}{\Huge\bf  Search for pair-produced }
\end{center}
\begin{center}{\Huge\bf  leptoquarks in e$^+$e$^-$ interactions}
\end{center}
\begin{center}{\Huge\bf  at $\sqrt{s}~\simeq~183$~GeV}
\end{center}
\bigskip
\bigskip
\bigskip
\begin{center}
{\bf \Large The OPAL Collaboration}
\end{center}
\bigskip
\bigskip\bigskip
\begin{center}{\Large\bf  Abstract}\end{center}
%
%
%
A search for pair-produced leptoquarks 
has been performed using a sample of $\mbox{e}^+\mbox{e}^-$ 
collision events collected by the OPAL detector at LEP at 
$\mbox{e}^+\mbox{e}^-$ centre-of-mass energies of about 183~GeV.
The data sample corresponds to an integrated luminosity of 
55.9~pb$^{-1}$.
The leptoquarks were assumed to be produced via couplings 
to the photon and the Z$^0$ and then to decay within a single fermion
generation.
No evidence for contributions from leptoquark pair production processes
was observed.
Lower limits on scalar and vector leptoquark masses are obtained. 
The existing limits are improved in the region of large decay branching 
ratio to quark-neutrino.
\bigskip
\begin{center}
{\large \bf to be submitted to \\
            European Physical Journal C } 
\end{center}
\bigskip
%
%
%
%
%
\end{titlepage}
\newpage
%
%
\begin{center}{\Large        The OPAL Collaboration
}\end{center}\bigskip
\begin{center}{
G.\thinspace Abbiendi$^{  2}$,
K.\thinspace Ackerstaff$^{  8}$,
G.\thinspace Alexander$^{ 23}$,
J.\thinspace Allison$^{ 16}$,
K.J.\thinspace Anderson$^{  9}$,
S.\thinspace Anderson$^{ 12}$,
S.\thinspace Arcelli$^{ 17}$,
S.\thinspace Asai$^{ 24}$,
S.F.\thinspace Ashby$^{  1}$,
D.\thinspace Axen$^{ 29}$
% ...
}\end{center}\bigskip
\bigskip
$^{  1}$School of Physics and Astronomy, University of Birmingham,
Birmingham B15 2TT, UK
\newline
$^{  2}$Dipartimento di Fisica dell' Universit\`a di Bologna and INFN,
I-40126 Bologna, Italy
\newline
% ...
$^{  8}$CERN, European Organisation for Particle Physics,
CH-1211 Geneva 23, Switzerland
\newline
$^{  9}$Enrico Fermi Institute and Department of Physics,
University of Chicago, Chicago IL 60637, USA
\newline
% ...
$^{ 12}$Indiana University, Department of Physics,
Swain Hall West 117, Bloomington IN 47405, USA
\newline
% ...
$^{ 16}$Department of Physics, Schuster Laboratory, The University,
Manchester M13 9PL, UK
\newline
$^{ 17}$Department of Physics, University of Maryland,
College Park, MD 20742, USA
\newline
% ...
$^{ 23}$Department of Physics and Astronomy, Tel Aviv University,
Tel Aviv 69978, Israel
\newline
$^{ 24}$International Centre for Elementary Particle Physics and
Department of Physics, University of Tokyo, Tokyo 113-0033, and
Kobe University, Kobe 657-8501, Japan
\newline
% ...
$^{ 29}$University of British Columbia, Department of Physics,
Vancouver BC V6T 1Z1, Canada
\newline
% ...
$^{  e}$ and University of Mining and Metallurgy, Cracow
\newline
% ...
\newline 
%
\newpage
%
\section{Introduction}
%
\indent
In the Standard Model (SM) quarks and leptons appear as formally 
independent components.
However, they show an apparent symmetry with respect to the
family and multiplet structure of the electroweak interactions.
It seems therefore natural that some theories beyond the SM~\cite{GUTS_COMP}
predict the existence of new bosonic fields, called leptoquarks (LQs),
mediating interactions between quarks and leptons.
The interactions of LQs with the known particles
are usually described by an effective Lagrangian that satisfies 
the requirement of baryon and lepton number conservation and respects the
SU(3)$_{\mathrm C}~\otimes$~SU(2)$_{\mathrm L}~\otimes$~U(1)$_{\mathrm Y}$ 
symmetry of the SM~\cite{LEPTOQ}.
This results in nine scalar ($S$) and nine vector ($V$) leptoquarks, 
grouped into 
weak isospin triplets ($S_1$ and $V_1$), doublets 
($S_{1/2}$, $\tilde{S}_{1/2}$, $V_{1/2}$ and $\tilde{V}_{1/2}$)
and singlets ($S_{0}$, $\tilde{S}_{0}$, $V_{0}$ and $\tilde{V}_{0}$)
\footnote{In this paper the notation used in~\cite{lowenergy} is adopted.
This is slightly different from the notation used in~\cite{LEPTOQ}.}.
They are shown in Tables~\ref{tab:scalarleptoquarks} 
and~\ref{tab:vectorleptoquarks}. 
%
\begin{table}[t]
\begin{center}
\begin{tabular}{|ccrcrcccccc|}
   \hline
   \hline
        & &       & &            & &       & &          & &          \\
~~LQ~~& & $I_3$ & & $Q_{e.m.}$ & & decay & & coupling $\lambda_{L,R}$ & &
                                                                    $\beta$  \\
        & &       & &            & &       & &          & &          \\
   \hline
        & &       & &            & &       & &          & &          \\
        & &       & &        & &$e^{-}_{L} u_{L}$ &:& $\lambda_{LS_{0}}$& &  \\
$S_0$   & &   0   & &$-1/3$  & &$e^{-}_{R} u_{R}$ &:& $\lambda_{RS_{0}}$& &
                     $\frac{\lambda_{LS_{0}}^{2}+\lambda_{RS_{0}}^{2}}
                           {2\lambda_{LS_{0}}^{2}+\lambda_{RS_{0}}^{2}}$ \\
        & &       & &      & &$\nu_{e} d_{L}$   &:& $-\lambda_{LS_{0}}$ & & \\
        & &       & &            & &                  & &           & &   \\
   \hline
              & &   & &            & &       & &          & &             \\
$\tilde{S}_{0}$& & 0 & &$-4/3$& &$e^{-}_{R} d_{R}$&:&$\lambda_{R\tilde{S}_{0}}$
                                                                        & & 1 \\
        & &       & &            & &       & &          & &          \\
   \hline
        & &       & &            & &       & &          & &          \\
       &  &  1 & &2/3   & & $\nu_{e} u_{L}$&:&$\sqrt{2}\lambda_{LS_{1}}$& & 0 \\
$S_1$  &  &  0 & &$-1/3$& &
                     $\left\{ \begin{array}{r} \nu_{e} d_{L}\\
                                               e^{-}_{L} u_{L}
                              \end{array} \right.$                 &
                            $ \begin{array}{r} : \\ : \end{array}$ &
                             $ \begin{array}{r} -\lambda_{LS_{1}} \\ 
                                                -\lambda_{LS_{1}}
                              \end{array} $
                            & & 1/2  \\
       &  &$-1$ & &$-4/3$ & & $e^{-}_{L} d_{L}$&:&$-\sqrt{2}\lambda_{LS_{1}}$
                                                                         & & 1\\
        & &       & &            & &       & &          & &          \\
   \hline
        & &       & &            & &       & &          & &          \\
        & &$1/2$ & &$-2/3$ & &
                     $\left\{ \begin{array}{r} \nu_{e} \overline{u}_{L} \\
                                               e^{-}_{R} \overline{d}_{R}
                              \end{array} \right.$         &
                     $\begin{array}{r}  : \\ : \end{array}$ &
                     $\begin{array}{r}  \lambda_{LS_{1/2}} \\ 
                                        -\lambda_{RS_{1/2}} \end{array}$ & &
                      $\frac{\lambda_{RS_{1/2}}^{2}}
                            {\lambda_{LS_{1/2}}^{2}+\lambda_{RS_{1/2}}^{2}}$ \\
$S_{1/2}$& &     & &             & &       & &          & &         \\
        & &$-1/2$& &$-5/3$& &
                     $\left\{ \begin{array}{r} e^{-}_{L} \overline{u}_{L} \\
                                               e^{-}_{R} \overline{u}_{R}
                      \end{array} \right.$  &
                     $\begin{array}{r}  : \\ : \end{array}$ &
                     $\begin{array}{r}  \lambda_{LS_{1/2}} \\ 
                                        \lambda_{RS_{1/2}}
                      \end{array}$ & & 1  \\
        & &       & &            & &       & &          & &          \\
   \hline
        & &       & &            & &       & &          & &          \\
        & &$1/2$ & &$1/3$ & &
                     $ \nu_{e} \overline{d}_{L}$ & : &
                                    $\lambda_{L\tilde{S}_{1/2}}$ & & 0\\
$\tilde{S}_{1/2}$& &     & &   & &   & &   & &  \\
        & &$-1/2$& &$-2/3$& &
                   $ e^{-}_{L} \overline{d}_{L}$ & : & 
                                     $\lambda_{L\tilde{S}_{1/2}}$ & & 1\\
        & &       & &            & &       & &          & &          \\
   \hline
   \hline
\end{tabular}
\end{center}
\caption{\sl {Quantum numbers and couplings for scalar leptoquarks.
         $Q_{e.m.}$ is the electric charge in units of $e$,
         $I_3$ the third component of the weak isospin
         and $\beta$ denotes the branching ratio of decay to a charged
         lepton and a quark. 
         Under the assumption of coupling within
         a single generation the same table must be repeated for second and
         third generation with the obvious substitutions $e\rightarrow
         \mu,\tau$, $u\rightarrow c,t$ and $d\rightarrow s,b$.}}
\label{tab:scalarleptoquarks}
\end{table} 
%
%
\begin{table}[t]
\begin{center}
\begin{tabular}{|ccrcrcccccc|}
   \hline
   \hline
        & &       & &            & &       & &          & &          \\
~~LQ~~& & $I_3$ & & $Q_{e.m.}$ & & decay & & coupling $\lambda_{L,R}$ & &
                                                                    $\beta$  \\
        & &       & &            & &       & &          & &          \\
   \hline
        & &       & &            & &       & &          & &          \\
    & &       & & & &$e^{-}_{L}\overline{d}_{R}$ &:& $\lambda_{LV_{0}}$& &  \\
$V_0$ & & 0 & &$-2/3$  & &$e^{-}_{R}\overline{d}_{L}$ &:& $\lambda_{RV_{0}}$& &
                     $\frac{\lambda_{LV_{0}}^{2}+\lambda_{RV_{0}}^{2}}
                           {2\lambda_{LV_{0}}^{2}+\lambda_{RV_{0}}^{2}}$ \\
    & &  & &      & &$\nu_{e}\overline{u}_{R}$ &:& $\lambda_{LV_{0}}$ & & \\
        & &       & &            & &                  & &           & &   \\
   \hline
              & &   & &            & &       & &          & &             \\
$\tilde{V}_{0}$& & 0 & &$-5/3$& &$e^{-}_{R}\overline{u}_{L}$
                                             &:&$\lambda_{R\tilde{V}_{0}}$
                                                                        & & 1 \\
        & &       & &            & &       & &          & &          \\
   \hline
        & &       & &            & &       & &          & &          \\
   &  &  1 & &1/3& & $\nu_{e}\overline{d}_{R}$&:&$\sqrt{2}\lambda_{LV_{1}}$
                                                                       & &0 \\
$V_1$  &  &  0 & &$-2/3$& &
                     $\left\{ \begin{array}{r} \nu_{e}\overline{u}_{R}\\
                                               e^{-}_{L}\overline{d}_{R}
                              \end{array} \right.$                 &
                            $ \begin{array}{r} : \\ : \end{array}$ &
                             $ \begin{array}{r} \lambda_{LV_{1}} \\ 
                                                -\lambda_{LV_{1}}
                              \end{array} $
                            & & 1/2  \\
  &  &$-1$& &$-5/3$ & & $e^{-}_{L}\overline{u}_{R}$&:&$\sqrt{2}\lambda_{LV_{1}}$
                                                                         & & 1\\
        & &       & &            & &       & &          & &          \\
   \hline
        & &       & &            & &       & &          & &          \\
        & &$1/2$ & &$-1/3$ & &
                     $\left\{ \begin{array}{r} \nu_{e} d_{R} \\
                                               e^{-}_{R} u_{L}
                              \end{array} \right.$         &
                     $\begin{array}{r}  : \\ : \end{array}$ &
                     $\begin{array}{r}  \lambda_{LV_{1/2}} \\ 
                                        \lambda_{RV_{1/2}} \end{array}$ & &
                      $\frac{\lambda_{RV_{1/2}}^{2}}
                            {\lambda_{LV_{1/2}}^{2}+\lambda_{RV_{1/2}}^{2}}$ \\
$V_{1/2}$& &     & &             & &       & &          & &         \\
        & &$-1/2$& &$-4/3$& &
                     $\left\{ \begin{array}{r} e^{-}_{L} d_{R} \\
                                               e^{-}_{R} d_{L}
                      \end{array} \right.$  &
                     $\begin{array}{r}  : \\ : \end{array}$ &
                     $\begin{array}{r}  \lambda_{LV_{1/2}} \\ 
                                        \lambda_{RV_{1/2}}
                      \end{array}$ & & 1  \\
        & &       & &            & &       & &          & &          \\
   \hline
        & &       & &            & &       & &          & &          \\
        & &$1/2$ & &$2/3$ & &
                     $ \nu_{e} u_{R}$ & : &
                                    $\lambda_{L\tilde{V}_{1/2}}$ & & 0\\
$\tilde{V}_{1/2}$& &     & &   & &   & &   & &  \\
        & &$-1/2$& &$-1/3$& &
                   $ e^{-}_{L} u_{R}$ & : & 
                                     $\lambda_{L\tilde{V}_{1/2}}$ & & 1\\
        & &       & &            & &       & &          & &          \\
   \hline
   \hline
\end{tabular}
\end{center}
\caption{\sl {Quantum numbers and couplings for vector leptoquarks.
         $Q_{e.m.}$ is the electric charge in units of $e$,
         $I_3$ the third component of the weak isospin
         and $\beta$ denotes the branching ratio of decay to a charged
         lepton and a quark.
         Under the assumption of coupling within
         a single generation the same table must be repeated for second and
         third generation with the obvious substitutions $e\rightarrow
         \mu,\tau$, $u\rightarrow c,t$ and $d\rightarrow s,b$.}}
\label{tab:vectorleptoquarks}
\end{table} 
Under these assumptions, only the mass and the couplings to right-handed
and/or left-handed fermions, denoted by $\lambda_{\mathrm{R}}$ and
$\lambda_{\mathrm{L}}$, remain as free parameters, since
the couplings to the electroweak gauge bosons are completely determined by 
the electric charge and the third component of the weak isospin.
Moreover, in order not to contradict the existing
indirect constraints on the leptoquark
masses and couplings coming from low
energy data such as rare decays~\cite{LEPTOQ,lowenergy}, 
the requirement that a given LQ couples to just one family of
fermions is imposed. \newline

Several experimental results constrain the existence of leptoquarks.
Searches for events with LQ single production, where a first generation LQ 
could be formed 
as a resonance
between an electron and a quark, were performed by
the ZEUS and H1 experiments at the $ep$ collider HERA~\cite{ZEUS_H1} 
and by the LEP experiments~\cite{delphisingle}. In $e^{+}e^{-}$ collisions
the quark comes from a resolved photon emitted by one 
of the LEP beams. As the production process directly involves a
LQ--lepton--quark interaction, limits on the LQ mass, M$_{\mathrm{LQ}}$,
can be derived as a function
of the couplings, $\lambda$, to fermions.
Leptoquark masses below about 80 GeV are excluded
for $\lambda$ values greater than a few $10^{-2}$ (about one order of
magnitude smaller than the electromagnetic coupling 
$\lambda_{\mathrm e}~\simeq$~0.3).
For $\lambda$~=~$\lambda_{\mathrm e}$, H1 excludes LQ
masses up to $275$~GeV.
Both LEP and FERMILAB experiments have searched for events with LQ pair
production~\cite{LEP1,CDF_D0}, setting limits
on M$_{\mathrm{LQ}}$ as a function of $\beta$, the branching ratio 
of decay into a charged lepton and a quark. On the contrary, these
limits do not depend on the couplings to fermions because the LQ pair
should be produced via coupling to the gauge bosons.
First generation scalar LQs with masses lower than about 200 GeV are excluded, 
assuming $\beta$~$\geq$~0.5, while for $\beta$~=~0 the mass limit 
is 79 GeV (D0).
Second generation LQs are excluded below 160~GeV, assuming $\beta$~=~0.5, and
below about 200~GeV, if $\beta$~=~1.
Third generation scalar leptoquarks 
with charge $|Q_{e.m.}|~=~\frac{2}{3}$ 
or $\frac{4}{3}$ and $\beta$~=~1 
are excluded for masses lower than 99 GeV by CDF, while 
D0 excludes third generation scalar leptoquarks with charge 
$|Q_{e.m.}|~=~\frac{1}{3}$ and $\beta$~=~0 below masses of 94 GeV.
In the same cases vector LQs belonging to the third generation 
are excluded for masses below 170~GeV (CDF) and 148~GeV (D0).
\newline

In principle, LQs of all three generations can be pair-produced 
in e$^+$e$^-$ collisions at LEP, 
by {\it s}-channel 
$\gamma$ or Z$^0$ exchange and, in the case of first generation 
LQs, by exchange of a quark in the {\it t}-channel~\cite{LEPTOQEE}.
Due to the existing upper limits on the couplings, $\lambda$, to fermions,
the {\it t}-channel contribution to the first generation production 
cross-section is negligible so that, for a given LQ state
\footnote{In this paper ``state" will be used to refer to a charge
eigenstate within a multiplet.
It will be denoted by S$_I$(Q$_{e.m.}$) or V$_I$(Q$_{e.m.}$), 
where the index I represents 
the weak isospin while the number in parenthesis is the electric charge
in units of e.}, 
the cross-section depends on the mass, the electric charge and third 
component of the weak isospin, but is independent 
of the $\lambda$ couplings.
On the other hand, for couplings smaller than ${\cal O}(10^{-5})$
the lifetime of leptoquarks would be sufficiently long to produce a
secondary decay vertex, clearly detatched from the primary vertex,
or even outside the detector. This topology is not considered here as
the charged tracks are required to come from the interaction vertex. 
To summarize, the present analysis covers the region of values
of the couplings to fermions from about $10^{-6}$
to $10^{-2}$.
The decay of a heavy LQ into a quark and a charged lepton
leads to final states characterized by an isolated energetic charged lepton, 
while for decays into a quark and a neutrino, the final state will have large
missing energy.
Therefore the following topologies can be considered for events 
that result from the
decay of a leptoquark-antileptoquark pair:\newline
{\bf Class A}: two hadronic jets and two neutrinos; it includes 
               the final states 
               $\nu_{e} \overline{\nu_{e}} u \overline{u}$,
               $\nu_{e} \overline{\nu_{e}} d \overline{d}$,
               $\nu_{\mu} \overline{\nu_{\mu}} c \overline{c}$,
               $\nu_{\mu} \overline{\nu_{\mu}} s \overline{s}$ and
               $\nu_{\tau} \overline{\nu_{\tau}} b \overline{b}$.
\newline
{\bf Class B}: two hadronic jets, one neutrino and one charged lepton of the 
first ($\nu_{e} e u d$) or second generation ($\nu_{\mu} \mu c s$).
In the hypothesis that each LQ couples to just one family of fermions,
the topology including two jets, one neutrino and one $\tau$ lepton is not
possible at LEP, since one of the two jets would have originated in
a top-quark.
\newline
{\bf Class C}: two hadronic jets and one pair of oppositely charged leptons
of the first or second generation, for example $e^{+} e^{-} u \overline{u}$
and $\mu^{+} \mu^{-} c \overline{c}$ respectively. 
\newline
{\bf Class D}: two hadronic jets and one pair of oppositely charged 
$\tau$ leptons, $\tau^{+} \tau^{-} b \overline{b}$. 
This case is considered separately from Class C because of the
different possible $\tau$ decays.  
\newline

In this paper a search is presented for pair-produced scalar
leptoquarks of all three generations performed with the OPAL detector.
Additionally, at this time, it was extended to vector leptoquarks of the 
first and second generation.
The pair production process gives the advantage, with respect to single
production by electron-quark interactions, that all states can 
be produced, included LQs coupling
exclusively to neutrinos. Therefore searches for this channel at LEP give 
the possibility to explore the region of large decay
branching ratio into quark-neutrino, where the FERMILAB experiments
have reduced sensitivity. Moreover this is the first search for pair 
produced vector LQs at LEP.  
The study is based on data recorded during the 1997 LEP run at
centre-of-mass energies, $\sqrt{s}$, between 181 and 184 GeV, corresponding 
to an integrated luminosity of 55.9~pb$^{-1}$.
The luminosity weighted average of $\sqrt{s}$ is 182.7~GeV.
Limits are derived under the assumption that 
only one state contributes to the cross-section. 
In the case of the third generation, mass limits will be given for scalar 
LQs which decay into a b-quark and either a $\tau$-lepton or a $\tau$-neutrino.
%
\section{The OPAL Detector}
%
\indent
The OPAL detector is described in detail in Ref.~\cite{opaltechnicalpaper}. 
It is a multi-purpose apparatus having nearly complete solid angle coverage 
\footnote{The right-handed coordinate system is defined so that $z$ is the
coordinate parallel to the e$^+$ and e$^-$ beams, with
positive direction along the e$^-$ beam; $r$ is the
coordinate normal to the beam axis, $\phi$ is the azimuthal angle with respect
to the positive direction of the {\it x}-axis (pointing towards the 
centre of LEP) and
$\theta$ is the polar angle with respect to +$z$.}.
The central detector consists of two layers of silicon micro-strip detectors 
\cite{simvtx} surrounding the beam-pipe and a system of tracking chambers 
inside a 0.435~T solenoidal magnetic field.
This system consists of a high-precision drift chamber, a large-volume 
jet chamber and a set of {\it z}-chambers measuring the track 
coordinates along the beam direction.
A lead-glass electromagnetic calorimeter 
is located outside the magnet coil and covers the full azimuthal range
in the polar angle range of \newline
$\mid\cos\theta\mid~<~0.984$.
It is divided into two regions: the barrel ($\mid\cos\theta\mid~<~0.82$)
and the endcaps ($\mid\cos\theta\mid~>~0.81$).
The magnet return yoke, divided into barrel and endcap sections along with
pole tips, is instrumented for hadron calorimetry in the region
$\mid\cos\theta\mid~<~0.99$.
Four layers of muon chambers cover the outside of the hadron calorimeter.
Close to the beam axis the forward calorimeter and gamma catcher 
together with the silicon-tungsten
luminometer \cite{bib-siw} complete the geometrical 
acceptance down to 33 mrad from the beam direction.
%
\section{Monte Carlo simulations}
%
Neglecting the {\it t}-channel contribution, the differential cross 
sections for the production of a pair of leptoquarks of mass 
M$_{\mathrm{LQ}}$ in $e^{+}e^{-}$ collisions at a 
centre-of-mass energy $\sqrt{s}$ are~\cite{LEPTOQEE}:
{\small
\begin{equation}
%
 \frac{d\sigma_{S}}{d\cos{\theta}}~=~
\frac{3\pi\alpha^2}{8s}
\left(1~-~4\mathrm{M}_{\mathrm{LQ}}^{2}/s\right)^{\frac{3}{2}}
sin^2\theta
                     \sum_{a=L,R} |k_a(s)|^2
\label{eq:totscalar}
\end{equation}}
{\small
\begin{equation}
\frac{d\sigma_{V}}{d\cos{\theta}}~=~
 \frac{3\pi\alpha^2}{8{\mathrm{M}}_{\mathrm{LQ}}^{2}}
  \left(1~-~4\mathrm{M}_{\mathrm{LQ}}^{2}/s\right)^{\frac{3}{2}}
  \left[1~+~\frac{1~-~3\left(1~-~4\mathrm{M}_{\mathrm{LQ}}^{2}/s\right)}{4}
   \sin^{2}\theta \right] 
    \sum_{a=L,R} |k_a(s)|^2
\label{eq:totvector}
\end{equation}}
\newline
for scalar and vector LQs respectively.~$\alpha$ is the electromagnetic coupling
and

{\small
\begin{equation}
 k_a(s)~=~-Q_{e.m.}~+~
      Q_{a}^{Z}({\mathrm e}) \frac{s}{s~-~M_{Z}^{2}~+~iM_{Z}\Gamma_{Z}}
       Q^{Z}({\mathrm{LQ}})
\end{equation}}
\newline
where $Q_{e.m.}$ is the LQ electric charge,
${\mathrm M}_{Z}$ and $\Gamma_Z$ are the mass and 
the width of the weak neutral current
gauge boson,
and 
the couplings are given by 
{\small
\begin{eqnarray}
     Q^{Z}(\mathrm{LQ})     &=& \frac{I_3~-~Q_{e.m.}\sin^2\theta_W}
                    {\cos{\theta_W}\sin{\theta_W}}   \nonumber \\
     Q^{Z}_{L}(\mathrm{e})  &=& \frac{-\frac{1}{2}~+~\sin^2\theta_W}
                                     {\cos{\theta_W}\sin{\theta_W}} 
                                      \label{eq:qzcouplings} \\
     Q^{Z}_{R}(\mathrm{e})  &=& \tan{\theta_W}  \nonumber
\end{eqnarray}}
\newline
In equations~\ref{eq:qzcouplings} $I_3$ is the third component of the LQ weak 
isospin and 
$\theta_W$ is the Weinberg angle. \newline

The Monte Carlo generator LQ2~\cite{LQ2} was used to simulate
leptoquark pair events.
Initial state QED radiation was included.
The leptoquarks are assumed to decay isotropically in their rest frame
and the hadronization of the
final state 
$\mathrm{q}\overline{\mathrm{q'}}$~pair was performed by JETSET~\cite{PYTHIA}.
Samples of 1000 signal events 
corresponding to different values of the leptoquark mass,
from M$_{{\mathrm LQ}}~=~50$~GeV to 
M$_{{\mathrm LQ}}~=~90$~GeV in steps of 5~GeV, were generated
for all the different decay topologies.
The full simulation of the response of the OPAL detector~\cite{GOPAL} 
was performed 
on the generated events.
Since they carry colour,
leptoquarks may hadronize before decaying,
if the couplings to fermions are small.
This possibility was taken into account by introducing a systematic
error on the detection efficiencies which was evaluated by using 
Monte Carlo samples in which the leptoquarks hadronized before decaying.
\newline

All relevant Standard Model background processes were studied using 
Monte Carlo generators. Two-fermion events 
(Z$^{0\ast}$/$\gamma^{\ast}$~$\rightarrow$~{\rm f\={f}}($\gamma$), 
{\rm f}~=~q,$\tau$)
were simulated with PYTHIA~\cite{PYTHIA} and KORALZ~\cite{KORALZ}.
The Monte Carlo programs HERWIG~\cite{HERWIG} and PHOJET~\cite{PHOJET} 
were used to generate two-photon
hadronic events. 
Other processes with four fermions in the final state, including 
W pair production, were simulated with grc4f \cite{GRACE} and
Vermaseren \cite{VERMASEREN}.
%
\section{Analysis}
%
Charged tracks used in the calculation of physical variables were required 
to have their origin close to the $\ee$ interaction point, to have at least 20 
measured space points in the jet chamber and at least 50\% of the hits 
geometrically expected.
The minimum transverse momentum of the track with respect to
the beam direction had to be greater than 120 MeV.
Electromagnetic clusters were required to have an energy of at 
least 100 MeV in the barrel and 
250 MeV in the endcaps.
Endcap clusters were also required 
to contain at least two adjacent 
lead glass blocks.
Clusters in the hadron calorimeter were rejected if their energy was
smaller than 0.6~GeV (2~GeV in the hadron poletips, i.e. for 
$\mid\cos\theta\mid~>~0.91$).
To avoid double counting, calculations of experimental quantities 
such as visible energy, transverse 
momentum, etc., were performed following the method explained in~\cite{MT_TN}.
\newline

All the different topologies of signal events 
(classes {\bf A} to {\bf D}) are characterized 
by large charged 
multiplicities and large energy depositions due to the hadronization 
of the $\mathrm{q}\overline{\mathrm{q'}}$~pair.
In this analysis no attempt was made to identify the flavour of 
the quarks.
Electron and muon identification, required in the selection of events of 
classes {\bf B} and {\bf C}, was performed 
by making a logical ``OR" of different 
standard algorithms~\cite{VDOPP}.
The electron identification is based on the energy match between a track
and the associated cluster in the electromagnetic calorimeter 
and uses a minimum of
subdetectors in order to give high efficiency, while the muon
identification requires a minimum number of associated hits in either
the muon chambers or hadron calorimeter strips.
The energy of identified electrons was taken from the energy of the 
electromagnetic calorimeter cluster associated to the identified 
electron track.
%
\subsection{The {\it jet~jet}~{\boldmath $\nu$~$\nu$} channel (class A)}
\begin{figure}[h]
\vspace*{-1cm}
\centerline{\epsfxsize=15.0cm\epsffile{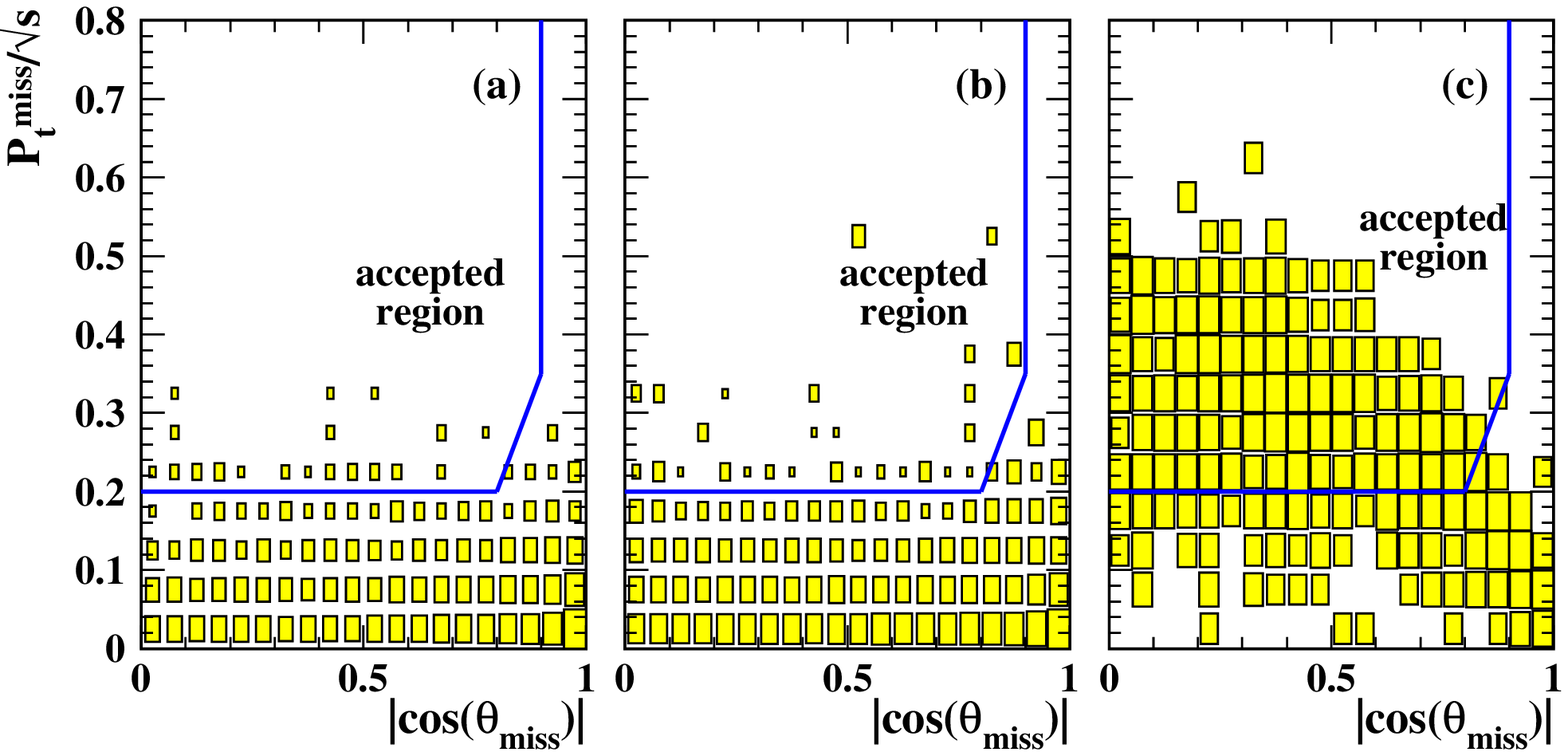}}
\vspace*{-7.8cm}
\caption[1]
{
\newline
Class {\bf A}, the {\it jet~jet}~$\nu~\nu$ channel: The distribution of events 
in the plane 
(p$_{\mathrm{t}}^{\mathrm{miss}}/\sqrt{s}$)~vs.
                                     $\mid$cos($\theta_{\mathrm{miss}}$)$\mid$,
shown after the preselection and cut A1, 
for the data {\bf (a)}, the simulated background {\bf (b)} and simulated 
scalar LQ
signal events of class~{\bf A} with M$_{\mathrm{LQ}}$~=~85~GeV {\bf (c)}.
The area of each box is proportional to the logarithm of the number 
of events falling within a 
two dimensional bin and is normalized with respect to the total 
content of each histogram separately.
}
\label{selea1}
\end{figure}

%
\indent
Signal events of class {\bf A} are characterized by an acoplanar pair of
hadronic 
jets and missing energy.
A few preselection requirements were applied to the data.
To reduce beam-wall and beam-gas interactions, 
the fraction of charged tracks that satisfied the quality criteria given above
was required to be greater than 0.2. A similar requirement was
applied to the non-associated electromagnetic clusters.
Both the number of accepted charged tracks
and the 
number of accepted non-associated electromagnetic clusters 
had to exceed four.
Finally, the total visible energy, E$_{\mathrm{vis}}$, was required to be
greater than 0.25$\sqrt{s}$ and smaller than 1.25$\sqrt{s}$.
After the preselection, the following cuts were applied to the data:
\begin{itemize}
\item[{\bf A1)}]
{The events contained no identified charged electron or muon 
of energy greater than 0.12$\sqrt{s}$.}
\item[{\bf A2)}]
{A two-dimensional cut was made in the plane 
(p$_{\mathrm{t}}^{\mathrm{miss}}$/$\sqrt{s}$)~vs.
$\mid\cos(\theta_{\mathrm{miss}})\mid$, 
where 
p$_{\mathrm{t}}^{\mathrm{miss}}$ is the missing transverse momentum 
with respect to the {\it z}-axis and
cos($\theta_{\mathrm{miss}}$) the cosine of the angle between
the missing momentum and the {\it z}-axis.
The cut is shown in Figure~\ref{selea1}.}
\item[{\bf A3)}]
{The events were reconstructed into two jets using the Durham
algorithm~\cite{DURHAM}: the sum of the energies of the two jets, 
E$_{\mathrm{jets}}$,
had to be such that 
$0.25\sqrt{s}<\mathrm{E}_{\mathrm{jets}}<0.75\sqrt{s}$.}
\item[{\bf A4)}]
{The jets were required to be acolinear by asking that
cos($\theta_{\mathrm{jj}}$)~$>$~--0.1, where
$\theta_{\mathrm{jj}}$ is the angle between the two jet directions.} 
\end{itemize}
Cuts A1 and A2 are useful, in particular, in rejecting two-photon events 
and greatly reducing 
Z$^{0*}$/$\gamma^*$~$\rightarrow$~{\rm f\={f}}($\gamma$) 
background.
Cuts A3 and A4 completely reject 
Z$^{0*}$/$\gamma^*$~$\rightarrow$~{\rm f\={f}}($\gamma$)
events, and reduce four-fermion background.\newline

\small
\begin{table}[htb]
\centering
\begin{tabular}{|c||c||c||c|c|c||c|c|}
\hline
& {\bf Data}    &  {\bf Background}            & 4-fermions  &
$\gamma \gamma$ & Z$^{0*}$/$\gamma^*$~$\rightarrow$~{\rm f\={f}}($\gamma$) & 
{\bf $\epsilon_{S}^{A}$~(\%)} & {\bf $\epsilon_{V}^{A}$~(\%)} \\ 
\hline
\hline
(A1)     &  16281 & 14913   &   685    &   9308   &  4920   & 90.0 & 88.4 \\ 
\hline
(A2)     &     87 &  80.9   &    52.2  &   1.6    &  27.1   & 59.4 & 58.7 \\ 
\hline
(A3)     &     53 &    46.8 &    42.1  &   0.9    &   3.8   & 58.5 & 56.9 \\ 
\hline
(A4)     &      4 &     3.6 &     2.7  &   0.8    &  $<$0.1 & 41.7 & 39.8 \\ 
\hline
%
\end{tabular}
\caption{\sl{
The remaining numbers of events after each cut of selection A
for various background processes are compared with the data.
The background is normalised to an integrated 
luminosity of 55.9~pb$^{-1}$.
The last two columns report the signal efficiencies, in percent,
for events with M$_{\mathrm{LQ}}$~=~85~GeV, for scalar and vector LQs
respectively.
}}
\label{tab:nevA}
\end{table}
\normalsize
Table~\ref{tab:nevA} shows the number of events after each cut, 
compared with the number of background events as predicted
from Monte Carlo samples, and the efficiencies for signal
events corresponding to M$_{\mathrm{LQ}}$~=~85~GeV.
Four events in the data survive the selections, in good agreement with
the number of expected background events from Standard Model processes,
3.57~$\pm$~0.83~({\it stat.}).
%
\subsection{The {\it jet~jet}~{\it l}
                {\boldmath$^{\pm}$~$\nu$} channel (class B)}
\begin{figure}[h]
\vspace*{-2.5cm}
\centerline{\epsfxsize=15.0cm\epsffile{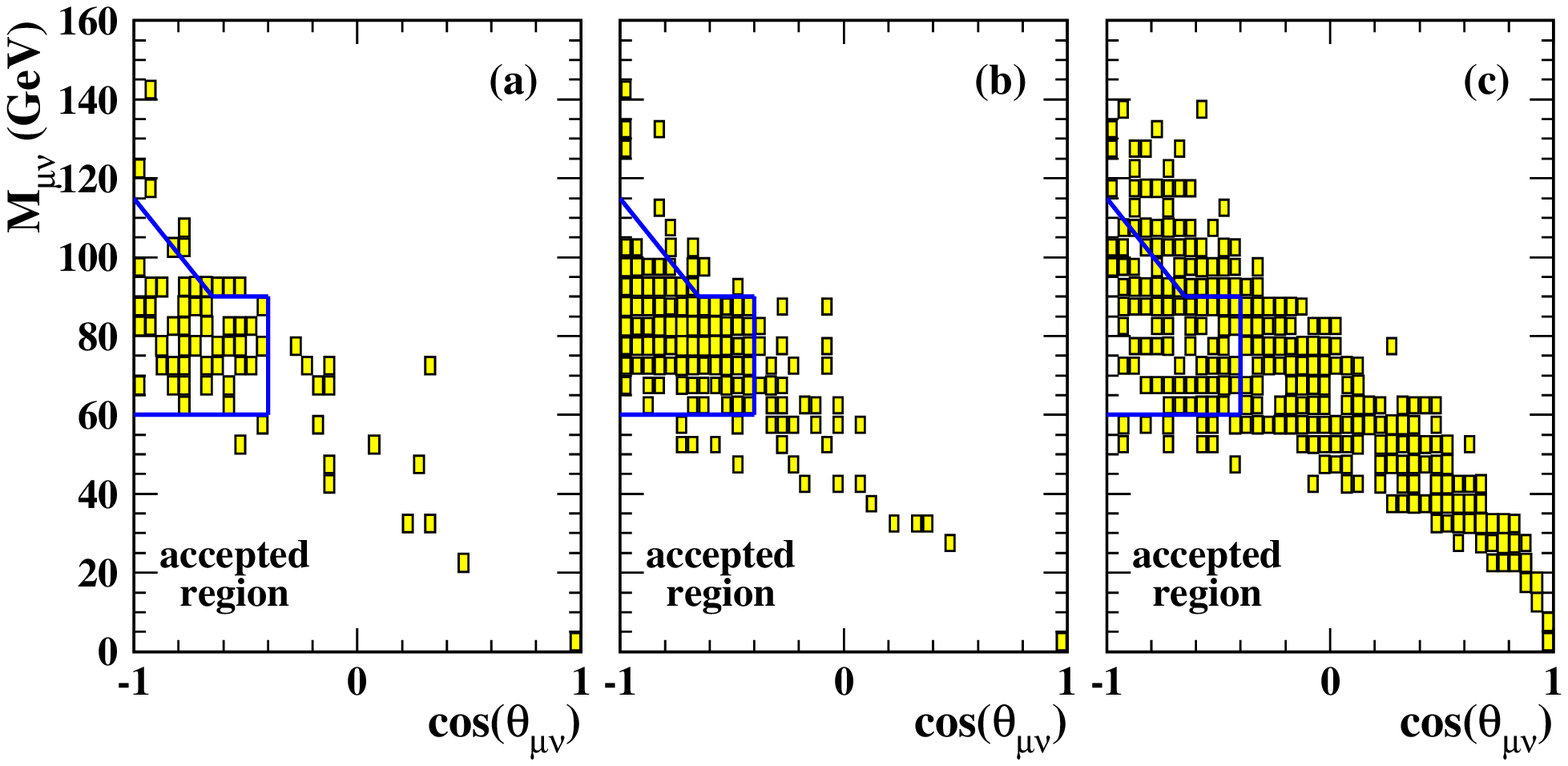}}
\vspace*{-7.8cm} 
\caption[1]
{
\newline
Class {\bf B}, the {\it jet~jet}~$l~\nu$ channel: The distribution 
of events in the plane
M$_{\mathrm{\mu\nu}}$~vs.~cos($\theta_{\mathrm{\mu\nu}}$),
shown after the preselection and cuts B1~--~B3,
for the data {\bf (a)}, the simulated background {\bf (b)} and simulated 
scalar LQ
signal events of class~B, second generation, with M$_{\mathrm{LQ}}$~=~85~GeV
{\bf (c)}.
The area of each box is proportional to the logarithm of the number 
of events falling within a 
two dimensional bin and is normalized with respect to 
the total content of each histogram
separately.}
\label{seleb1}
\end{figure}

\begin{figure}[h]
\centerline{\epsfxsize=15.0cm\epsffile{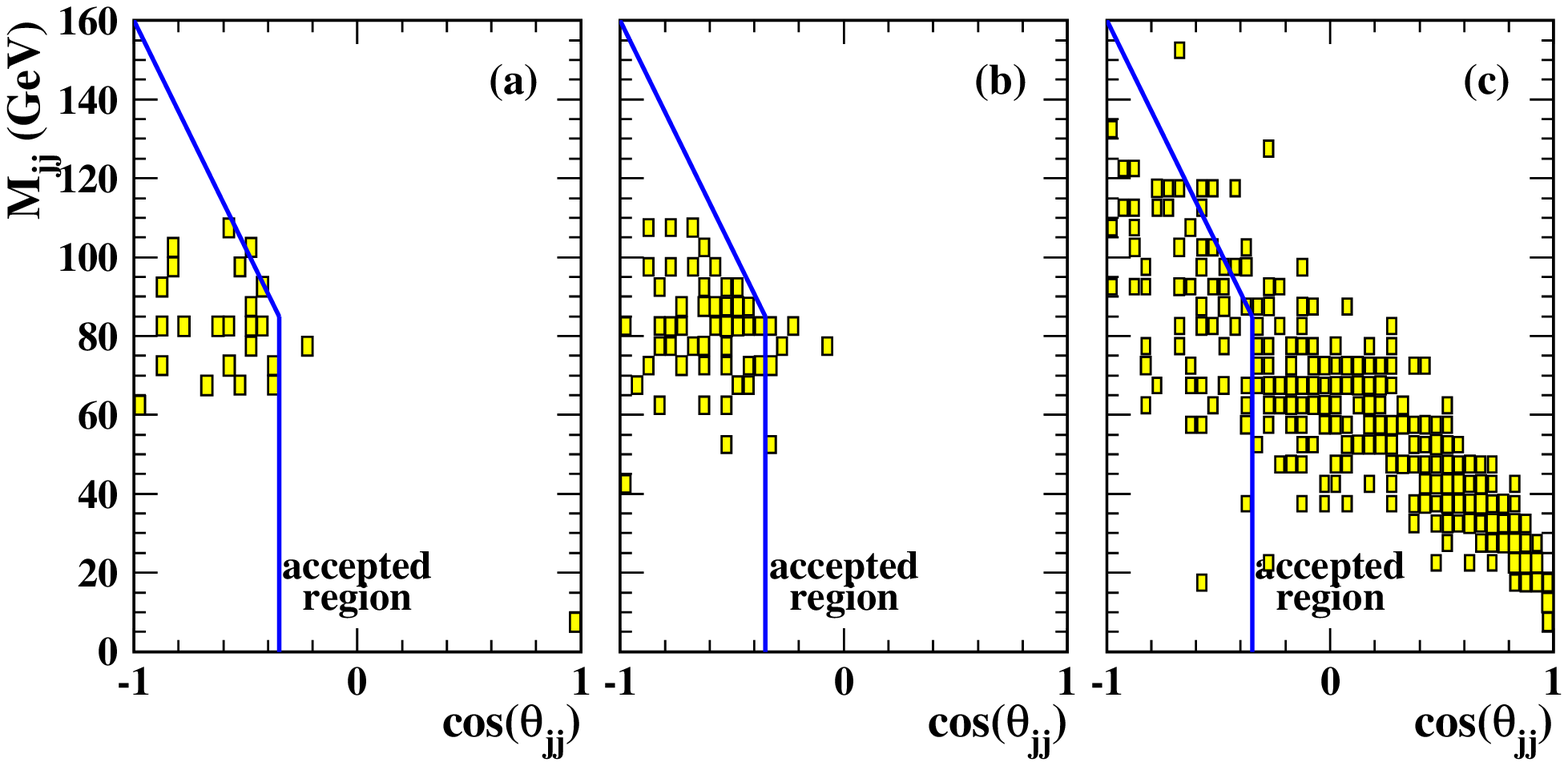}}
\vspace*{-7.8cm} 
\caption[1]
{
\newline
Class {\bf B}, the {\it jet~jet}~$l~\nu$ channel: The distribution of 
events in the plane
M$_{\mathrm{jj}}$~vs.~cos($\theta_{\mathrm{jj}}$),
displayed after the preselection and cuts B1~--~B4, 
for the data {\bf (a)}, the total simulated background {\bf (b)} 
and simulated scalar LQ
signal events of class~B with M$_{\mathrm{LQ}}$~=~85~GeV {\bf (c)}.
The area of each box is proportional to the logarithm of the number 
of events falling within a 
two dimensional bin and is normalized with respect to 
the total content of each histogram
separately.}
\label{seleb2}
\end{figure}

%
\indent
After the application of the same preselection cuts described in section 4.1,
the selection of signal events of class {\bf B} proceeded as follows:
\begin{itemize}
\item[{\bf B1)}]
{The event was required to contain at least one identified electron or
muon with an energy of at least 0.1$\sqrt{s}$.}
\item[{\bf B2)}]
{No charged tracks and at most one (no) electromagnetic
cluster within a cone of half-aperture 15$^{\circ}$ around 
the most energetic electron (muon).}
\item[{\bf B3)}]
{The missing transverse momentum
had to be greater 
than 0.09$\sqrt{s}$.}
\item[{\bf B4)}]
{A 2-dimensional cut in the plane
M$_{\mathrm{l\nu}}$~vs.~cos($\theta_{\mathrm{l\nu}}$)
was applied as shown in Figure~\ref{seleb1}, where $\theta_{\mathrm{l\nu}}$ 
is the angle between the momentum vector of the most energetic lepton 
and the missing momentum vector,
while M$_{\mathrm{l\nu}}$ is their invariant mass.}
\item[{\bf B5)}]
{After having removed the most energetic lepton, the event was
forced into two jets using the Durham algorithm.
The sum of the energies of the two jets, E$_{\mathrm{jets}}$,
had to be such that $0.25\sqrt{s}<\mathrm{E}_{\mathrm{jets}}<0.75\sqrt{s}$.
Moreover, a 2-dimensional cut in the plane 
cos($\theta_{\mathrm{jj}}$)~vs.~M$_{\mathrm{jj}}$
was applied as shown in Figure~\ref{seleb2}, where $\theta_{\mathrm{jj}}$ 
is the angle between the two jet directions and M$_{\mathrm{jj}}$ is the 
invariant mass of the two-jet system.}
\end{itemize}
\small
\begin{table}[htb]
\centering
\begin{tabular}{|c||c||c||c|c|c||c|c|}
\hline
& {\bf Data}    &  {\bf Background}            & 4-fermions  &
$\gamma \gamma$ & Z$^{0*}$/$\gamma^*$~$\rightarrow$~{\rm f\={f}}($\gamma$) &
{\bf $\epsilon_{S}^{B}$~(\%)} & {\bf $\epsilon_{V}^{B}$~(\%)}\\ 
\hline
\hline
(B1) & 1110{\bf /}304 & 1106{\bf /}321 & 277{\bf /}151 & 28.4{\bf /}2.5 
             & 801{\bf /}169 & 89.1{\bf /}87.7 & 90.7{\bf /}90.1 \\ 
\hline
(B2) &  196{\bf /}109 & 179{\bf /}101 & 128{\bf /}99.6 & 20.8{\bf /}0.
             &  30.2{\bf /}1.4 & 79.8{\bf /}80.3   & 81.7{\bf /}79.9 \\ 
\hline
(B3) &  108{\bf /}95  &  98.4{\bf /}85.9  & 94.6{\bf /}85.5 & 0.8{\bf /}0.
             & 3.0{\bf /}0.4 & 73.5{\bf /}72.3 & 75.2{\bf /}73.6 \\ 
\hline
(B4) &   23{\bf /}29  & 19.4{\bf /}20.7   & 17.1{\bf /}20.3 
          & 0.8{\bf /}0. & 1.5{\bf /}0.4 & 59.8{\bf /}58.8 & 60.4{\bf /}61.7 \\ 
\hline
(B5) &    1{\bf /}3   &  2.2{\bf /}2.9   &  2.0{\bf /}2.6 
          & 0.{\bf /}0. &  0.2{\bf /}0.3  & 43.1{\bf /}49.1 & 43.6{\bf /}51.5\\ 
\hline
%
\end{tabular}
\caption{\sl{
The remaining numbers of events after each cut of selection B
for various background processes are compared with the data.
The background is normalised to an integrated luminosity of 55.9 pb$^{-1}$.
The last two columns report the signal efficiencies, in percent, for events with
${\mathrm{M_{LQ}}}$~=~85~GeV, for scalar and vector LQs respectively. 
When two numbers are separated by a slash, the first one refers to the
first generation and the second one to the second generation.
}}
\label{tab:nevB}
\end{table}
\normalsize
Two-photon events contribute negligibly to the background after selection B3.
Cuts B2 and B3 are particularly efficient in reducing 
Z$^{0*}$/$\gamma^*$~$\rightarrow$~{\rm f\={f}}($\gamma$) events.\newline

In Table \ref{tab:nevB} the numbers of events after each cut are shown,
compared with the numbers of predicted background events and the 
efficiencies for signal events corresponding to 
${\mathrm{M_{LQ}}}$~=~85~GeV. 
One event in the data is retained after the selection for the
first generation and three events for the second generation, 
in good agreement with an
expected background of 2.22~$\pm$~0.21~({\it stat.}) and 
2.84~$\pm$~0.18~({\it stat.}) events, respectively.
%
\subsection{The {\it jet~jet}~{\it l{\boldmath$^+$~l$^-$}} channel (class C)}
\begin{figure}
\vspace*{-2.5cm}
\hspace*{0.5cm}
\centerline{\epsfxsize=17.0cm\epsffile{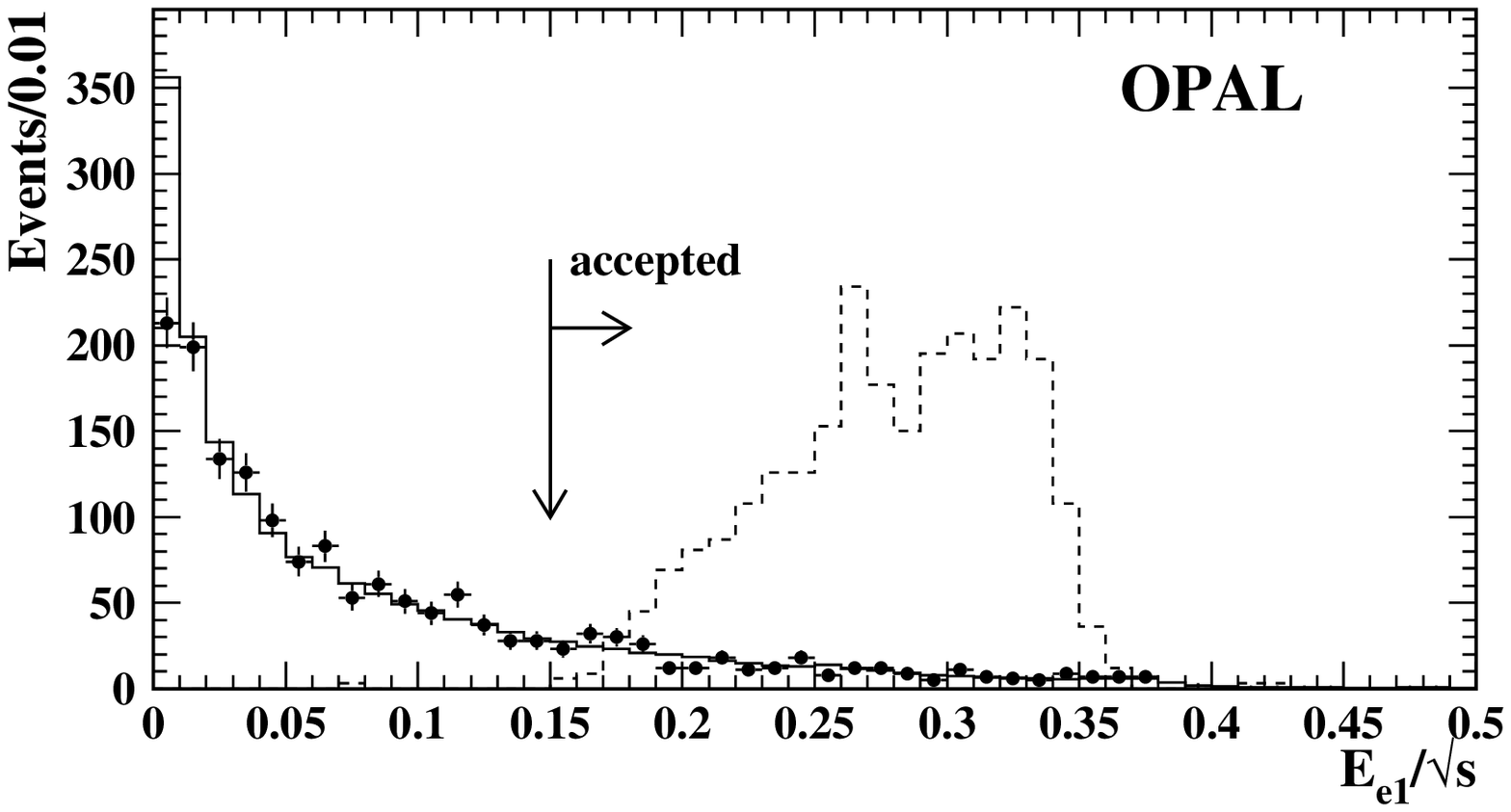}}
\vspace*{-9.8cm} 
\caption[1]
{
\newline
Class {\bf C}, the {\it jet~jet}~$l^{+}~l^{-}$ channel: The distribution 
of the energy of the most energetic electron for events 
containing an electron pair after the preselections, for the data (points), 
the simulated background (full line, normalised to the integrated luminosity
of the data) and a simulated signal (dashed line, arbitrary normalisation)
corresponding to ${\mathrm{M_{LQ}}}$~=~85~GeV. 
The arrow denotes the position of cut C1.}  
\label{selec1}
\end{figure}

%
\indent
Signal events of this type have small missing energy and are characterized 
by the presence of a pair of high energy charged leptons of the same
generation that tend to be
isolated.
The same preselections as for classes {\bf A} and {\bf B} were applied.
The following cuts were then applied to select events:
\begin{itemize}
\item[{\bf C1)}]
{The presence of at least one pair of oppositely-charged identified 
electrons or muons 
was required.
The most energetic leptons
of the same generation
and opposite in charge will be called the ``pair" in the following.
The energy of the most energetic lepton of the pair had to exceed 
0.15$\sqrt{s}$, while an energy of at least 0.1$\sqrt{s}$
was required for the other lepton.
For the first generation,
the distributions of the energy of the most energetic electron of the pair 
for data, expected background and signal events
corresponding to M$_{{\mathrm LQ}}$~=~85~GeV,
are shown in Figure~\ref{selec1}.}
\item[{\bf C2)}]
{No charged tracks and at most one (no) electromagnetic cluster within 
a cone of half-aperture 15$^{\circ}$ around the most energetic 
electron (muon). The same requirements, but considering a cone of
half-aperture 10$^{\circ}$, were applied for the less energetic lepton of
the pair.}
\item[{\bf C3)}]
{The two leptons of the pair were required to be acolinear: 
$\cos(\theta_{\mathrm{ll}})$~$>$~--0.85, where $\theta_{\mathrm{ll}}$
is the angle between the two leptons.}
\end{itemize}
\small
\begin{table}[htb]
\centering
\begin{tabular}{|c||c||c||c|c|c||c|c|}
\hline
& {\bf Data}    &  {\bf Background}            & 4-fermions &
$\gamma \gamma$ & Z$^{0*}$/$\gamma^*$~$\rightarrow$~{\rm f\={f}}($\gamma$)& 
 {\bf $\epsilon_{S}^{C}$~(\%)} & {\bf $\epsilon_{V}^{C}$~(\%)}\\ 
\hline
\hline
(C1)  &  97{\bf /}10 & 96.0{\bf /}7.4 & 19.9{\bf /}4.3 & 0.4{\bf /}0. 
                   &  75.7{\bf /}3.1   & 80.7{\bf /}80.3 & 79.0{\bf /}77.4 \\
\hline
(C2)  &   2{\bf /}2 & 2.8{\bf /}1.5 & 2.7{\bf /}1.5 & 0.1{\bf /}0.
                   & $<$0.1{\bf /}0.     & 67.2{\bf /}69.2 & 67.4{\bf /}67.3\\
\hline
(C3)  &   2{\bf /}2 & 1.6{\bf /}1.0 & 1.6{\bf /}1.0   & 0.{\bf /}0.   
                   & $<$0.1{\bf /}0.    & 63.7{\bf /}65.7 & 63.7{\bf /}63.9 \\
\hline
\end{tabular}
\caption[]{\sl{
The remaining numbers of events after each cut of selection C
for various background processes are compared with the data.
The background is normalised to an integrated luminosity of 55.9 pb$^{-1}$.
The last two columns report the signal efficiencies, in percent, for events with
${\mathrm{M_{LQ}}}$~=~85~GeV, for scalar and vector LQs respectively. 
When two numbers are separated by a slash, the first one refers to the
first generation and the second one to the second generation.
}}
\label{tab:nevC}
\end{table}
\normalsize
Background from two-photon events is removed after selection C1.
Cut C2 suppresses
Z$^{0*}$/$\gamma^*$~$\rightarrow$~{\rm f\={f}}($\gamma$) events.
The lepton acolinearity requirement C3 is useful in further reducing 
four-fermion background.\newline

The numbers of events after each cut, compared with the numbers of expected 
background events and the efficiencies for signal events corresponding
to ${\mathrm{M_{LQ}}}$~=~85~GeV, are shown in Table~\ref{tab:nevC}.
After all cuts two candidates survive the selection both for the 
first and the second generation; 
1.63~$\pm$~0.14~({\it stat.}) background events 
for the first generation and 0.98~$\pm$~0.15~({\it stat.}) for the second 
generation are expected.
%
\subsection{The {\it jet~jet}~{\boldmath $\tau^+$~$\tau^-$} channel (class D)}
\begin{figure}[h]
\vspace*{-1.5cm}
\hspace*{0.5cm}
\centerline{\epsfxsize=17.0cm\epsfysize=17.0cm\epsffile{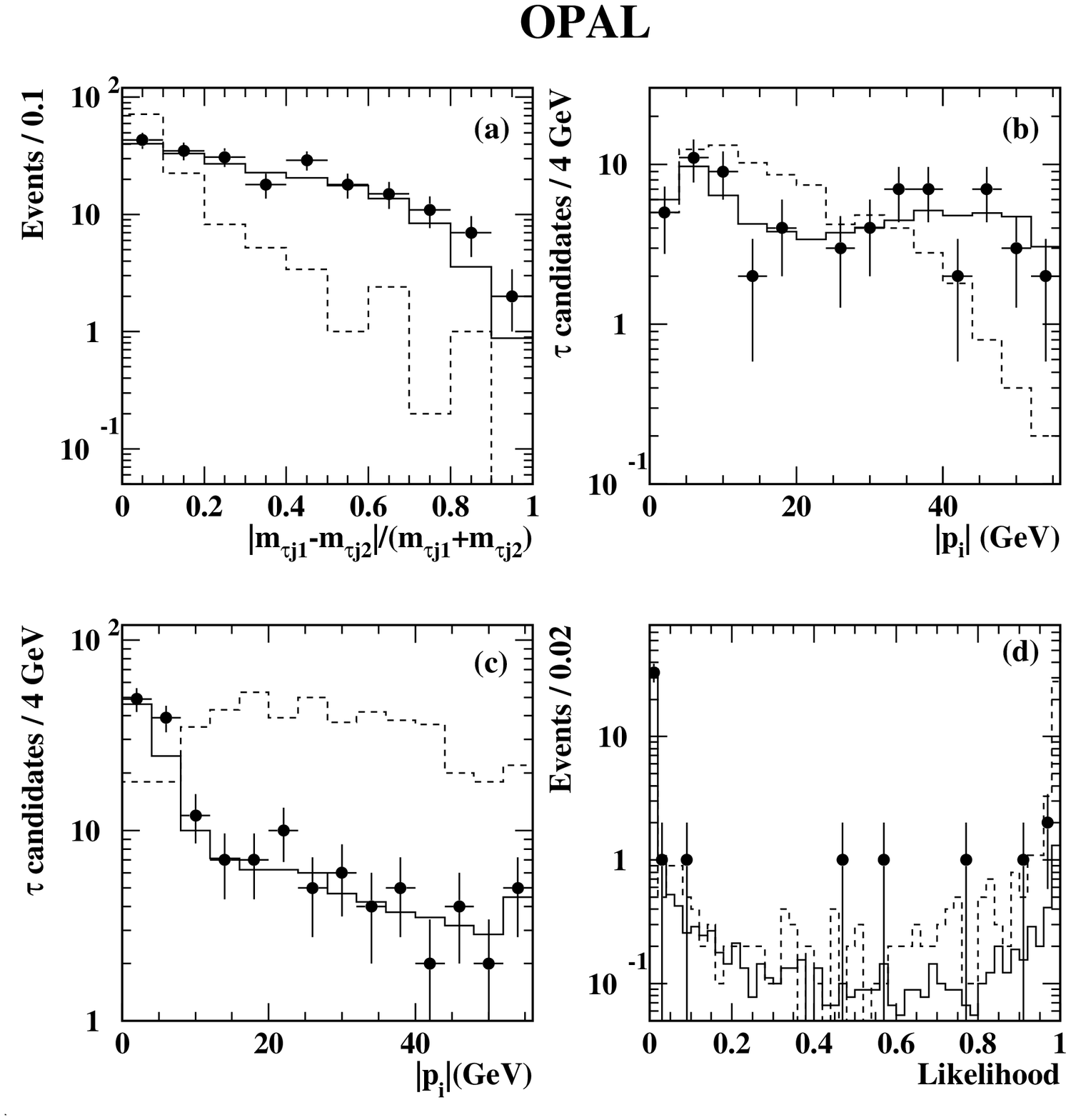}}
\vspace*{-1.0cm}
\caption[1]
{
 \newline
Class {\bf D}, the {\it jet~jet}~$\tau^{+}~\tau^{-}$ channel: Kinematic 
distributions 
for the data (points), estimated Standard Model background (full line,
normalised to the integrated luminosity of the data), and a simulated signal
(dashed line, arbitrary normalisation) corresponding to 
M$_{\mathrm{LQ}}$~=~85~GeV.
\newline
{\bf (a)}
 Distribution of the difference in invariant mass of the tau-jet
systems scaled by their sum after cut (D2).
\newline
Figures (b) and 
(c) demonstrate 
the difference in the distributions of the same likelihood input variable
for two different categories of $\tau$ candidate, after cut (D4):
{\bf (b)}
 The momentum of leptonic $\tau$ candidates.
{\bf (c)} 
The momentum of 1-prong hadronic $\tau$ candidates. 
{\bf (d)}
The likelihood distribution after cut (D6).
}
\label{fig:twotaufourjet}
\end{figure}

%
Signal events of this type are characterized by a pair of isolated
$\tau$-leptons and
a pair of energetic hadronic jets.
The background comes predominantly from 
Z$^{0*}$/$\gamma^{*}\rightarrow \mathrm{f}\bar{\mathrm{f}}(\gamma)$ 
and four-fermion processes.
The selection begins with the identification of $\tau$ lepton candidates, 
which is identical
to that in~\cite{smpaper}, using three specific algorithms 
to identify semileptonic and hadronic $\tau$-lepton decays.
On average, 2.3 $\tau$ candidates per signal event are thus identified.
The original $\tau$-lepton direction is approximated
by that of the visible decay products.
The following requirements are then imposed:
\begin{description}
\item[D1)]  
Events are required to contain at least nine charged tracks, and must have at 
least two $\tau$-lepton
candidates, including at least one pair whose members
each have electric charge $|q|=1$ and whose charges sum to zero.
Pairs not fulfilling these requirements are not considered further.

\item[D2)]
Events must have no more than 20 GeV of energy
deposited in the forward calorimeter, gamma-catcher, and 
silicon-tungsten luminometer; a missing momentum vector
satisfying $|\cos\theta_{\mathrm miss}| < 0.97$,
a total vector transverse momentum of at least 0.02~$\sqrt{s}$, and a scalar sum
of all track and cluster transverse momenta larger than 40~GeV.
  
\item[D3)]
Events must contain at least three jets, including single 
electrons and muons from $\tau$-lepton decay which are allowed 
to be recognised as low-multiplicity jets, 
reconstructed using the cone algorithm as in~\cite{smpaper}
and no energetic isolated photons.
An energetic isolated photon is defined as an electromagnetic cluster
with energy larger than 15~GeV and no track within a cone
of $30^\circ$ half-angle. 

\item[D4)]
Events must contain no track or cluster with 
energy exceeding $0.3\sqrt{s}$. 

\end{description}
For events surviving these requirements, 
the tracks and clusters not belonging to the $\tau$~pair
(henceforth referred to as the ``rest of the event'' or RoE),
are then split into two jets using the Durham algorithm.  Two pairing 
schemes between the two $\tau$ candidates and the jets are thus possible.
The invariant masses $m_{\tau j}$ of the two resulting $\tau$-jet systems 
within each pairing scheme are then calculated using only
the $\tau$-lepton and jet momentum directions and requiring energy and
momentum conservation. The pairing scheme 
exhibiting the lesser difference between $m_{\tau j1}$ and 
$m_{\tau j2}$ is then chosen. 
Then, in order for a $\tau$ candidate pair to be considered 
further, the following requirements on $m_{\tau j1}$ and 
$m_{\tau j2}$  are imposed, consistent
with the hypothesis of the decay of two heavy objects of identical mass:
\begin{description}
\item[D5)]
Both $m_{\tau j1}$ and $m_{\tau j2}$ must be
at least 30 GeV.
\item[D6)]
The difference in invariant masses must be no more than
30\% of their sum.
\end{description}
The distribution of 
$|m_{\tau j1}-m_{\tau j2}|/|m_{\tau j1}+m_{\tau j2}|$  is shown in
Fig.~\ref{fig:twotaufourjet}~(a) for the data,
the background events, and
for a signal sample with M$_{\mathrm{LQ}}$~=~85~GeV. 
The resolution on $m_{\tau j}$ is typically less than 7 GeV, except 
very close to the kinematic limit.

A likelihood method~\cite{rpvgauginos,rpvsfermions}
is then
applied to those events satisfying the above requirements, 
in order to select a final $\tau$ candidate pair for each
event from those surviving, and to suppress further the remaining background.
In each such event, for each remaining $\tau$ candidate pair and 
its associated hadronic RoE,
a joint discriminating variable, ${\cal L}$, is constructed 
using normalised reference distributions generated from Monte Carlo samples
of signal and background events.
The set of variables for
the reference distributions includes some which characterize each of the
two $\tau$-lepton candidates individually, some which describe their behavior
as a pair and some which characterize the RoE. For those variables
describing the $\tau$ candidates individually, 
separate reference distributions are generated for 
leptonic (electron or muon), 
hadronic 1-prong and hadronic 3-prong $\tau$ candidates, in order to exploit
the differences between the three categories.
Distributions of some of these input variables as well as that of 
${\cal L}$ are shown in 
Fig.~\ref{fig:twotaufourjet}~(b) to (d).
The $\tau$ candidate pair having the highest value of ${\cal L}$
is chosen as the definitive pair for each event, and the following
requirement is then made:

\begin{description}
\item[D7)] ${\cal L}>0.93$
\end{description}

Table~\ref{tautau_t0} shows the numbers of observed and expected
events after each requirement, along with the detection efficiency 
for a signal with M$_{\mathrm{LQ}}$~=~85~GeV.
Two events survive the selection while the background, predominantly
from four-fermion processes, is estimated to be
$2.07~\pm~0.15$~({\it stat.}) events.
\small
\begin{table}[htb]
\centering
\begin{tabular}{|c||c||c||c|c|c||c|}
\hline
& {\bf Data}    &  {\bf Background}            & 4-fermions  &
$\gamma \gamma$ & Z$^{0*}$/$\gamma^*$~$\rightarrow$~{\rm f\={f}}($\gamma$) &
{\bf $\epsilon_{S}^{D}$~(\%)}\\
\hline
\hline
(D1)     &1322   & 1070  & 150 & 791 & 129  & 59.4  \\
\hline
(D2)     & 209  &  191  & 129 & 3.0 & 59.1 & 57.8 \\
\hline
(D3)     & 198  & 181   & 127 & 2.9 & 51.1  & 57.5 \\
\hline
(D4)     & 149 & 139    & 100 & 2.2 & 36.3  & 56.2 \\
\hline
(D5)     &  51 & 55.5  & 47.5 & 0. & 8.0  & 52.7 \\
\hline
(D6)     & 41  & 44.8 & 38.1  & 0. & 6.7 & 50.2 \\
\hline
(D7)     & 2  & 2.1  & 2.1  & 0. & $<$0.02  & 32.6 \\
\hline
\end{tabular}
\caption{\sl{
The remaining numbers of events after each cut of selection D
for various background processes are compared with the data.
The background is normalised 
to an integrated luminosity of 55.9 pb$^{-1}$.
The last column reports the signal efficiency, in percent,
for events with M$_{{\mathrm LQ}}$~=~85~GeV for scalar LQs.}}
\label{tautau_t0}
\end{table}
\normalsize
%
%

\section{Results}
%
The detection efficiencies for the different topologies of signal events, 
as a function of the leptoquark mass M$_{\mathrm{LQ}}$, are listed in 
Table~\ref{effic183}.
\begin{table}[htb]
\centering
\begin{tabular}{|l|c||c|c|c|c|c|c|}
\hline
\multicolumn{2}{|c||}{M$_{{\mathrm LQ}}$ (GeV)} &50 & 60 & 70 & 80 & 85 & 90 \\
\hline
Signal topology & Generation&      &       &       &       &       &       \\
\hline
\hline
 Class A (scalar) & 1,2,3 
 & 14.3 & 23.9 & 31.4 & 37.0 & 41.7 & 42.8 \\
 Class A (vector) & 1,2   
 & 15.0 & 22.7 & 32.3 & 36.3 & 39.8 & 42.3 \\
\hline
 Class B (scalar) & 1 
 & 14.1 & 25.5 & 30.9 & 40.0 & 42.4 & 45.2 \\
 Class B (scalar) & 2 
 & 15.1 & 28.9 & 38.2 & 44.8 & 49.1 & 52.5 \\
 Class B (vector) & 1 
 & 13.9 & 23.0 & 35.5 & 41.7 & 43.6 & 46.7 \\
 Class B (vector) & 2 
 & 15.3 & 25.8 & 38.5 & 47.0 & 51.5 & 53.9 \\
\hline
 Class C (scalar) & 1 
 & 36.2 & 45.5 & 56.6 & 59.3 & 63.7 & 64.5 \\
 Class C (scalar) & 2 
 & 37.1 & 46.1 & 55.9 & 62.9 & 65.7 & 67.0 \\
 Class C (vector) & 1
 & 33.8 & 43.0 & 52.8 & 62.1 & 63.7 & 67.0 \\
 Class C (vector) & 2
 & 35.5 & 41.7 & 53.4 & 63.0 & 63.9 & 66.8 \\
\hline
 Class D (scalar) & 3
 & 22.3 & 27.4 & 29.6 & 32.0 & 32.6 & 32.8 \\
\hline
\end{tabular}
\caption[]{\sl{
The detection efficiencies for the various selections, in percent,
as a function of the leptoquark mass, M$_{{\mathrm LQ}}$.
}}
\label{effic183}
\end{table}

The systematic uncertainties in the number of signal events for 
the following sources were evaluated (they are quoted as relative \%):
\begin{itemize}
\item The errors due to signal Monte Carlo statistics and the
interpolation errors at an arbitrary point of M$_{\mathrm{LQ}}$
were estimated to be 5--10$\%$, depending on the signal topology.
\item The error associated with the electron and muon identification method
was evaluated using mixed events constructed by overlaying
Z$^{0*}$/$\gamma^{*}\rightarrow\mathrm{q}\overline{\mathrm {q}}$ events
with single hemispheres of
Z$^{0*}$/$\gamma^{*}\rightarrow\mathrm{l}^{+}\mathrm{l}^{-}$
$(\mathrm{l} = \mathrm{e}, \mu) $ events
at $\sqrt{s} \approx M_{Z^{0}}$~\cite{WW}.
Such events are topologically and kinematically analogous to 
q\={q}l$\nu$ events at $\sqrt{s}\approx 183$~GeV. This error was found to be
3.2$\%$ for electrons and 2.5$\%$ for muons.
The error for $\tau$ identification was evaluated to be
1.2$\%$~\cite{rpvsfermions}. 
\item The uncertainty introduced by the modelling of the variables
used in the selections was estimated to contribute with 5--15\%.
This was evaluated by displacing the cut values
by an amount corresponding
to the difference between the mean values of the data 
and background distributions.
The corresponding displacement in the detection efficiencies was taken
as a systematic error. The contributions from each single cut were added
in quadrature. 
%
\item The uncertainty in the flavour of the final state quarks
contributes 
with less than 6$\%$ for class {\bf A} and with less than 3$\%$ for  
class {\bf C}. 
This was evaluated by comparing the efficiencies 
corresponding to all the different possible flavours in a given decay channel,
characterized by the leptons in the final state (for example
$e^{+}e^{-}u\overline{u}$ and $e^{+}e^{-}d\overline{d}$ for class {\bf C}, 
first generation).
The value of the efficiency was taken to be the mean value and the largest
difference between the mean and the single contributions was taken as a
systematic error. 
In the case of classes {\bf B} and {\bf D} the flavours of the two quarks are
precisely determined by the hypothesis of coupling within a single
generation and (for class {\bf D}) the top-quark mass threshold and
thus no error is assigned.
\item The systematic error associated with the fragmentation of the 
leptoquark decay products was estimated to be about 10$\%$.
This was evaluated by comparing the signal efficiencies with the 
efficiencies obtained by applying the same selection cuts to signal events
in which the two LQs fragment before they decay.
\item The error on the integrated luminosity was 0.5$\%$.
\end{itemize}
The errors were considered to be independent and added in quadrature.
A total systematic error of 10--30$\%$ was estimated on the number
of expected background events, from the following sources: 
Monte Carlo statistics (10--20$\%$), modelling of cut variables (10--25$\%$), 
lepton identification (1.2--3.2$\%$) 
and integrated luminosity (0.5$\%$).\newline

No evidence of leptoquark production was observed in the data.
Upper limits at 95$\%$ confidence level (C.L.) 
on the LQ pair production 
cross-section $\sigma(\ee\rightarrow\mathrm{LQ\overline{LQ}})$ were
computed from the observed numbers of events,
the signal detection efficiencies and the number of expected background events, 
using the procedure described in~\cite{PDG}.
The uncertainties both on the efficiencies and on the background
were incorporated in the upper limits by numerical integration, 
assuming Gaussian distributions,
as suggested in~\cite{COUSINS}.
To obtain the limit at a given point in the plane $\beta$~vs~M$_{\mathrm{LQ}}$,
the independent analyses, corresponding to different classes of events, 
were combined by considering the total number of expected signal events.
For example, for the first and the second generation, this number is given by
\begin{equation}
      N^{tot}_{exp}~=~\sigma~\cdot~\int{{\cal L}}~\cdot~
        \left[\beta^{2}\epsilon^{C} + 2\beta(1-\beta)\epsilon^{B} +
              (1-\beta)^{2}\epsilon^{A}\right]     
      \nonumber    
\end{equation}
where $\sigma$ is the production cross-section,  
$\displaystyle\int{\small{\cal{L}}}$~ is 
the integrated luminosity of the data and $\epsilon^{A,B,C}$
are the signal detection efficiencies for classes A, B and C respectively. 
\newline

\begin{figure} 
\vspace*{-3.0cm}
\hspace*{0.5cm}
\centerline{\epsfxsize=17.5cm\epsffile{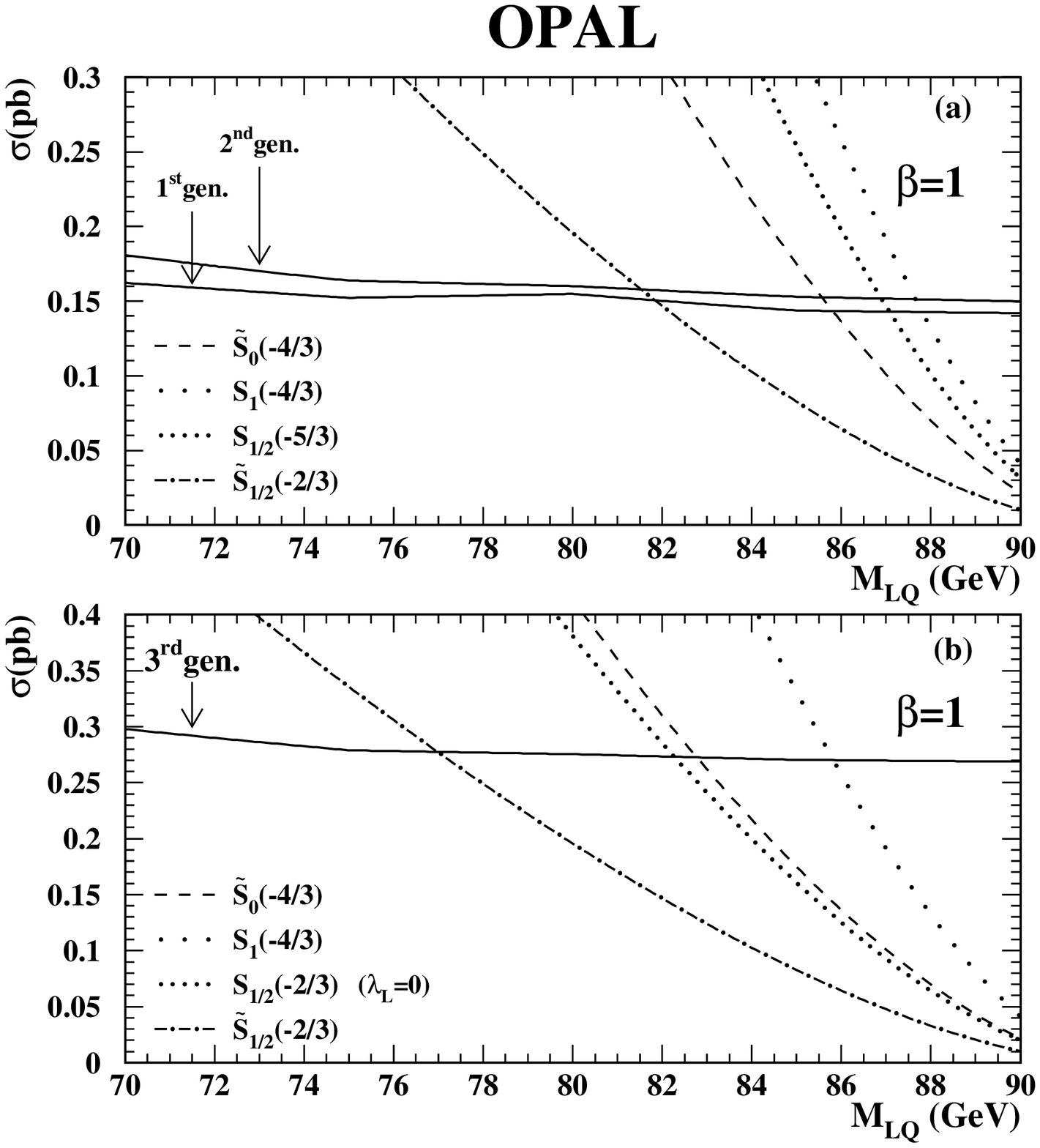}}
\caption[1]
{
\newline
{\bf (a)}
The 95$\%$ C.L. upper limits of the production cross-section for scalar LQs
of the first and second generation with $\beta$~=~1 (full lines)
as a function of the LQ mass, compared with the theoretical production 
cross-sections. 
{\bf (b)}
Same as (a) for third generation scalar LQs.
}
\label{limit1}
\end{figure}

\begin{figure}
\vspace*{-3.0cm}
\hspace*{0.5cm}
\centerline{\epsfxsize=17.5cm\epsffile{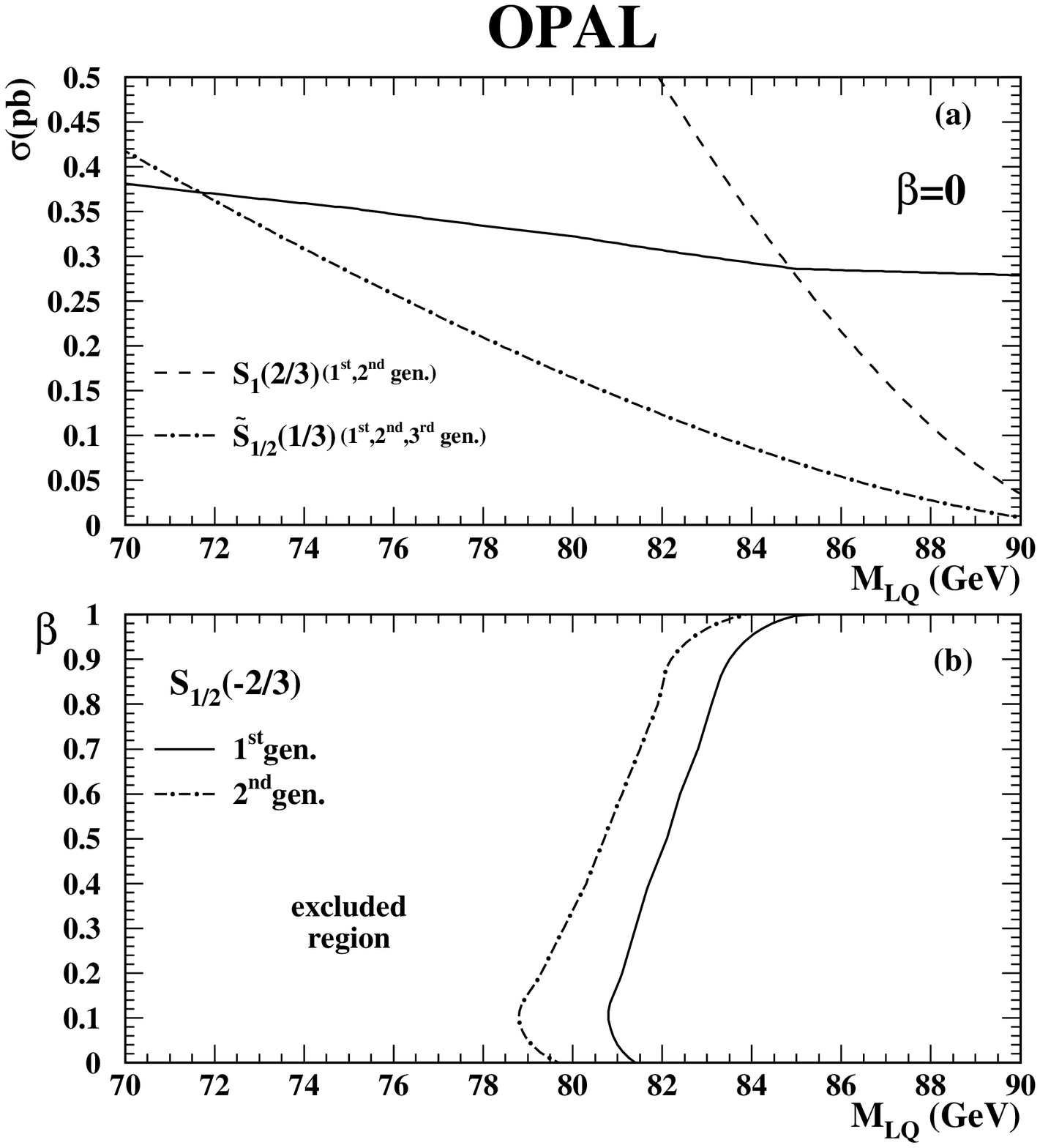}}
\caption[1]
{
\newline
{\bf (a)}
The 95$\%$ C.L. upper limit of the production cross-section for scalar LQs
with $\beta$~=~0 (full line) as a function of the LQ mass, compared with
the theoretical production cross-sections.
{\bf (b)}
The region of the plane $\beta$~vs~M$_{\mathrm{LQ}}$ excluded at 
the 95$\%$ C.L.
for the state $S_{1/2}(-2/3)$ of the first and second generation.
}
\label{limit2}
\end{figure}

\begin{figure} 
\vspace*{-3.0cm}
\hspace*{0.5cm}
\centerline{\epsfxsize=17.5cm\epsffile{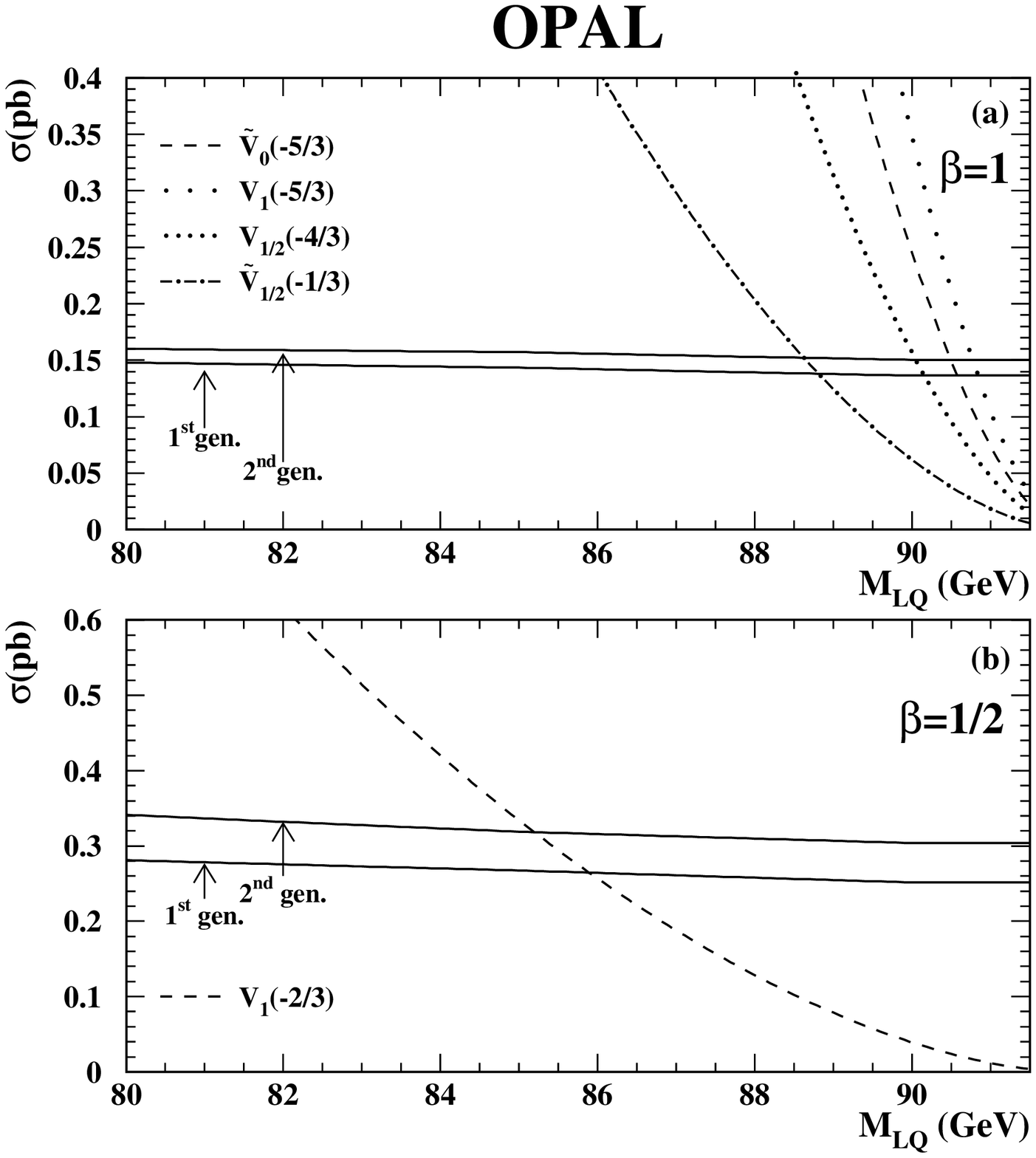}}
\caption[1]
{
\newline
{\bf (a)}
The 95$\%$ C.L. upper limits of the production cross-section for vector LQs
of the first and second generation with $\beta$~=~1~ (full lines)
as a function of the LQ mass, compared with the theoretical 
production cross-sections. 
{\bf (b)}
Same as (a) for vector LQs of the first and second generation 
with $\beta$~=~1/2.
}
\label{limit1vec}
\end{figure}

\begin{figure}
\vspace*{-3.0cm}
\hspace*{0.5cm}
\centerline{\epsfxsize=17.5cm\epsffile{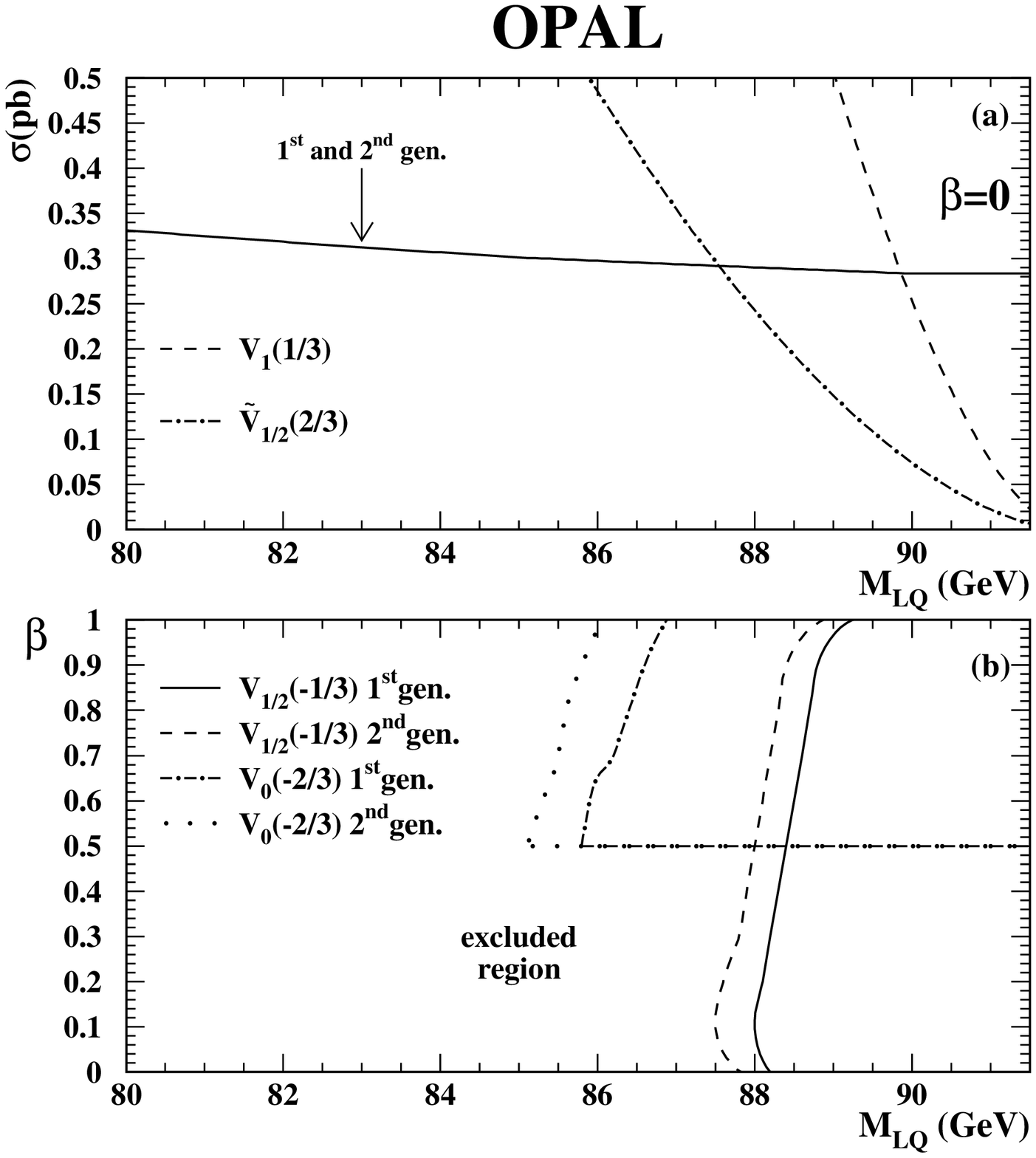}}
\caption[1]
{
\newline
{\bf (a)}
The 95$\%$ C.L. upper limit of the production cross-section for vector LQs
with $\beta$~=~0 (full line) as a function of the LQ mass, compared with
the theoretical production cross-sections.
{\bf (b)}
The region of the plane $\beta$~vs~M$_{\mathrm{LQ}}$ excluded at 
the 95$\%$ C.L.
for the states $V_{0}(-2/3)$ and $V_{1/2}(-1/3)$ 
of the first and second generation.
}
\label{limit2vec}
\end{figure}

Regions of the plane $\beta$~vs~M$_{\mathrm{LQ}}$ excluded at the 95\%~C.L. 
were determined by comparing the upper limits on the
production cross-sections with the total cross-sections computed 
by integration of Equations~\ref{eq:totscalar} and~\ref{eq:totvector}. 
In the case of the third generation, limits have been evaluated for
scalar states 
which decay only either into a $\tau$-lepton and a b-quark, 
or into a $\nu_{\tau}$ and a b-quark.
For the third generation state $S_{1/2}(-2/3)$ only $\beta = 1$  
can be considered since
below the top-quark threshold only the right-handed coupling is possible.
Figures~\ref{limit1}~(a),(b) and \ref{limit2}~(a) show the upper
limits of the production cross-sections as functions of the LQ mass, for
scalar LQs with $\beta$~=~0 and $\beta$~=~1.
The limit on $\tilde{S}_{1/2}(1/3)$ is valid for all three
generations but the same is not true for the limit on $S_1(2/3)$, since the
third generation state, under the hypothesis of diagonal coupling, would
decay to a top-quark and a $\nu_{\tau}$.
Figures~\ref{limit1vec}~(a),(b) and \ref{limit2vec}~(a) show the upper
limits of the production cross-sections as functions of the LQ mass, for
vector LQs with $\beta$~=~1, $\beta$~=~0.5 and $\beta$~=~0 respectively.
The regions excluded in the plane $\beta$~vs~M$_{\mathrm{LQ}}$ of the states
$S_{1/2}(-2/3)$, $V_{1/2}(-1/3)$ and $V_{0}(-2/3)$, whose $\beta$ can range 
from 0 to 1 (0.5 to 1 for $V_{0}$) according to
the relative weights of the left and right $\lambda$ couplings,
are reported in Figures~\ref{limit2}~(b) and~\ref{limit2vec}~(b) 
for the first and the second 
generation.
The mass limits obtained from the present analysis are summarized in 
Table~\ref{tab:limits}.
Because of their much smaller cross-sections
this search is not sensitive to the production of the states $S_0(-1/3)$ 
and $S_1(-1/3)$, so previous OPAL limits are quoted.
{\small
\begin{table}[ht]
\begin{center}
\hspace*{-30pt}
\begin{tabular}{|cccc|ccccc|}
   \hline
   \hline
 &            & &         &               & &               & &               \\
LQ &$Q_{e.m.}$  & & $\beta$ & 1$^{st}$~gen. & & 2$^{nd}$~gen. & & 
                                                              3$^{rd}$~gen. \\
 &            & &         &               & &               & &               \\
\hline
\hline
$S_0$ & $-1/3$ & & [0.5,1] &  44.2($\ast$)& &44.2($\ast$)& &  --            \\
\hline
$\tilde{S}_0$  & $-4/3$     & &    1    & 85.8  & & 85.5  & & 82.7          \\
\hline
             & $-2/3$     & & [0,1] &80.8($\ast\ast$)& &78.8($\ast\ast$)& 
                                                   &82.2($\ast\ast\ast$)    \\
$S_{1/2}$      &            & &         &       & &       & &               \\
               & $-5/3$     & &    1    & 87.0  & & 86.8  & & --            \\
\hline
               & $+1/3$     & &    0    & 71.6          & & 71.6 & & 71.6  \\
$\tilde{S}_{1/2}$ &         & &         &       & &       & &               \\
               & $-2/3$     & &    1    & 81.8  & & 81.5  & & 76.9          \\
\hline
               & $+2/3$     & &    0    & 84.8  & & 84.8 & & --            \\
         &            & &         &       & &      & &               \\
$S_1$    & $-1/3$     & &   0.5   & 44.2($\ast$)& &44.2($\ast$)& & --      \\
         &            & &         &       & &      & &               \\
               & $-4/3$     & &    1    & 87.8  & &87.6  & & 85.8          \\
\hline 
\hline               
 $V_0$  & $-2/3$     & & [0.5,1] & 85.8($\ast\ast$) & &85.1($\ast\ast$) & &
                                                                  --        \\
\hline
$\tilde{V}_0$  & $-5/3$     & &    1    & 90.5  & & 90.4  & & --          \\
\hline
               & $-1/3$     & & [0,1] &88.0($\ast\ast$)& &87.5($\ast\ast$)& 
                                                              & --       \\
$V_{1/2}$      &            & &         &       & &       & &               \\
               & $-4/3$     & &    1    & 90.1  & & 90.0  & & --            \\
\hline
           & $+2/3$     & &    0    & 87.5            & & 87.5 & & --  \\
$\tilde{V}_{1/2}$ &         & &         &       & &       & &               \\
               & $-1/3$     & &    1    & 88.8  & & 88.6  & & --          \\
\hline
               & $+1/3$     & &    0    & 89.8  & & 89.8 & & --            \\
               &            & &         &       & &      & &               \\
$V_1$          & $-2/3$     & &   0.5   & 85.8  & & 85.1 & & --            \\
               &            & &         &       & &      & &               \\
               & $-5/3$     & &    1    & 90.8  & &90.7  & & --          \\
\hline
\hline
\end{tabular}
\end{center}
\caption{\sl{Lower mass limits, in GeV, for scalar and vector leptoquarks as 
             obtained from the present analysis.\newline
             ($\ast$) LEP1 limits from OPAL. \newline
             ($\ast\ast$) Minimum allowed value for M$_{{\mathrm LQ}}$.
                          See Figures 7(b) and 9(b) for limits as functions
                          of $\beta$. \newline 
($\ast\ast\ast$) This limit is valid for $\beta$~=~1 (i.e. $\lambda_{L}$~=~0 
             below the top-quark threshold).}}
\label{tab:limits}
\end{table}
}
\section{Conclusions}
%
A data sample collected with the OPAL detector at $\sqrt{s}$~=~183~GeV,
corresponding to an integrated luminosity of 55.9~pb$^{-1}$, 
was analysed to search for events with pair production
of scalar and vector leptoquarks.
The search included scalar states of all three generations and
vector states of the first and second generation only. In the case of vector 
leptoquarks this is the first pair production search at LEP.
The present analysis covers the region of small values of the couplings,
$\lambda$, to fermions (about from 10$^{-6}$ to 10$^{-2}$).
No significant excess with respect to Standard Model predictions was found 
in the data. 
Lower mass limits for leptoquarks were set, under the assumption that 
only one leptoquark contributes to the cross-section.
The results improve previous LEP lower limits on scalar leptoquark masses
by 25--40 GeV, depending on the leptoquark quantum numbers.
With respect to the existing limits from CDF and D0 experiments, mass lower
limits are improved by about 10 GeV, for first and second generation, 
in the region of small values of the branching ratio of decay to a charged
lepton and a quark.
%
 
\newpage\newpage

\newpage


\newpage
%
\begin{center}
{\Large \bf OPAL}
\end{center}
\vspace*{-0.5cm}
\begin{center}
{\Large \bf OPAL}
\end{center}
\vspace*{1cm}
\newpage
\vspace*{0.4cm}
\begin{center}
{\Large \bf OPAL}
\end{center}
\vspace*{-1.5cm}
\vspace*{-2.5cm}
\newpage
\vspace*{2cm}
\newpage
\newpage
\newpage
\newpage
\end{document}